# Superstatistics in Random Matrix Theory

A.Y. Abul-Magd

Department of Basic Sciences, Faculty of Engineering, Sinai University,
El-Arish, Egypt

**Abstract:** Random matrix theory (RMT) provides a successful model for quantum systems, whose classical counterpart has a chaotic dynamics. It is based on two assumptions: (1) matrix-element independence, and (2) base invariance. Last decade witnessed several attempts to extend RMT to describe quantum systems with mixed regular-chaotic dynamics. Most of the proposed generalizations keep the first assumption and violate the second. Recently, several authors presented other versions of the theory that keep base invariance on the expense of allowing correlations between matrix elements. This is achieved by starting from non-extensive entropies rather than the standard Shannon entropy, or following the basic prescription of the recently suggested concept of superstatistics. The latter concept was introduced as a generalization of equilibrium thermodynamics to describe non-equilibrium systems by allowing the temperature to fluctuate. We here review the superstatistical generalizations of RMT and illustrate their value by calculating the nearest-neighbor-spacing distributions and comparing the results of calculation with experiments on billiards modeling systems in transition from order to chaos.

## Introduction

In classical mechanics, integrable Hamiltonian dynamics is characterized by the existence of as many conserved quantities as degrees of freedom. Each trajectory in the corresponding phase space evolves on an invariant hyper-torus [1]. In contrast, chaotic systems are ergodic; almost all orbits fill the energy shell in a uniform way. Physical systems with integrable and fully chaotic dynamics are exceptional. A typical Hamiltonian systems show a mixed phase space in which regions of regular motion and chaotic dynamics coexist. These systems are known as mixed systems. Their dynamical behavior is by no means universal. If we perturb an integrable system, most of the periodic orbits on tori with rational frequencies disappear. However, some of these orbits persist. Elliptic periodic orbits appear surrounded by islands. They correspond to librational motions around these periodic orbits and reflect their stability. The Kolmogorov-Arnold (KAM) theorem establishes the stability with respect to small perturbations of invariant tori with a sufficiently incommensurate frequency vector. When the perturbation increases, numerical simulations show that more and more tori are destroyed. For large enough perturbations, there are locally no tori in the considered region of phase-space. The break-up of invariant tori leads to a loss of stability of the system, to chaos. Different scenarios of transition to chaos in dynamical systems have been considered. There are three main scenarios of transition to global chaos in finite-dimensional (non-extended) dynamical systems: via the cascade of period-doubling bifurcations, Lorenz system-like transition via Hopf and Shil'nikov bifurcations, and the transition to chaos via intermittences [2-4]. It is natural to expect that there could be other (presumably many more) such scenaria in extended (infinite-dimensional) dynamical systems.

In quantum mechanics, the specification of a wave function is always related to a certain basis. In integrable systems eigenbasis of the Hamiltonian is known in principle. In this basis, each eigenfunction has just one component that obviously indicates the absence of complexity. In the nearly ordered regime, mixing of quantum states belonging to adjacent levels can be ignored and the energy levels are uncorrelated. The level-spacing distribution function obeys the Poissonian, $\exp(-s)$, where $s$ is the energy spacing between adjacent levels normalized by the mean level spacing. On the other hand, the eigenfunctions a Hamiltonian with a chaotic classical limit is unknown in principle. In other words, there is no special basis to express the eigenstates of a chaotic system. If we try to express the wave functions of a chaotic system in terms of a given basis, their components become on average uniformly distributed over the whole basis. They are also extended in all other bases. For example, Berry [5] conjectured that the wavefunctions of chaotic quantum systems can be represented as a formal sum over elementary solutions of the Laplace equation in which real and imaginary parts of coefficients are independent identically-distributed Gaussian random variables with zero mean and variance computed from the normalization. Bohigas et al. [6] put forward a conjecture (strongly supported by accumulated numerical evidence) that the spectral statistics of chaotic systems follow random-matrix theory (RMT) [7, 8]. This theory models a chaotic system by an ensemble of random Hamiltonian matrices H that belong to one of the three universal classes, orthogonal, unitary and symplectic. The theory is based on two main assumptions: the matrix elements are independent identically-distributed random variables, and their distribution is invariant under unitary distributions. These lead to a Gaussian probability density distribution for the matrix elements. The Gaussian distribution is also obtained by maximizing the Shannon entropy under constraints of normalization and existence of the expectation value of Tr($H^{\dagger}H$), where Tr denotes the trace and $H^{\dagger}$ stands for the Hermitian conjugate of H. The statistical information about the eigenvalues and/or eigenvectors of the matrix can be obtained by integrating out all the undesired variables from distribution of the matrix elements. This theory predicts a universal form of the spectral correlation functions determined solely by some global symmetries of the system (time-reversal invariance and value of the spin). Time-reversal-invariant quantum systems are represented by a Gaussian orthogonal ensemble (GOE) of random matrices when the system has rotational symmetry and by a Gaussian symplectic ensemble (GSE) otherwise. Chaotic systems without time reversal invariance are represented by the Gaussian unitary ensemble (GUE). Among several measures representing spectral correlations, the nearest-neighbor level-spacing distribution function $P(s)$ has been extensively studied so far. For GOE, the level spacing distribution function in the chaotic phase is approximated by the Wigner-Dyson distribution, namely,

$$P(s) = \frac{\pi}{2} s e^{-\frac{\pi}{4}s^2}.$$

Analogous expressions are available for GUE and GSE [7].

The assumptions that lead to RMT do not apply for mixed systems. The Hamiltonian of a typical mixed system can be described as a random matrix with some (or all) of its elements as randomly distributed. Here the distributions of various matrix elements need not be the same, may or may not be correlated and some of them can be non-random too. This is a difficult route to follow. So far in the literature, there is no rigorous statistical description for the transition from integrability to chaos. There have been several proposals for phenomenological random matrix theories that interpolate between the Wigner-Dyson RMT and banded RM ensemble with the (almost) Poissonian level statistics, in which the level spacing distribution is given by the Poisson distribution

$$P(s) = e^{-s}$$

The standard route of the derivation is to sacrifice basis invariance but keep matrix-element independence. The first work in this direction is due to Rosenzweig and Porter [10]. They model the Hamiltonian of the mixed system by a superposition diagonal matrix of random elements having the same variance and a matrix drawn from a GOE. Therefore, the variances of the diagonal elements total Hamiltonian are different from those of the off-diagonal ones, unlike the GOE Hamiltonian in which the variances of diagonal elements are twice of the off-diagonal ones. Hussein and Pato [11] used the maximum entropy principle to construct such ensembles by imposing additional constraints. Ensembles of band random matrices whose entries are equal to zero outside a band of width $b$ along the principal diagonal have often been used to model mixed systems [12]

Another route for generalizing RMT is to conserve base invariance but allow for correlation of matrix elements. This has been achieved by maximizing non-extensive entropies subject to the constraint of fixed expectation value of Tr($H^\dagger H$) [13]. Recently, an equivalent approach is presented in [14], which is based on the method of superstatistics (statistics of a statistics) proposed by Beck and Cohen [15]. This formalism has been applied successfully to a wide variety of physical problems, e.g., in [15]. In thermostatics, superstatistics arises as weighted averages of ordinary statistics (the Boltzmann factor) due to fluctuations of one or more intensive parameter (e.g. the inverse temperature). Its application to RMT assumes the spectrum of a mixed system as made up of many smaller cells that are temporarily in a chaotic phase. Each cell is large enough to obey the statistical requirements of RMT but has a different distribution parameter η associated with it, according to a probability density $f(\eta)$. Consequently, the superstatistical random-matrix ensemble that describes the mixed system is a mixture of Gaussian ensembles with a statistical weight $f(\eta)$. Therefore one can evaluate any statistic for the superstatistical ensemble by simply integrating the corresponding statistic for the conventional Gaussian ensemble.

## Beck and Cohen's superstatistics

Consider a complex system in a nonequilibrium stationary state. Such a system will be, in general, inhomogeneous in both space and time. Effectively, it may be thought to consist of many spatial cells, in each of which there may be a different value of some relevant intensive parameter. For instance, a system with Hamiltonian $H$ at thermal equilibrium is well represented by a canonical ensemble. The distribution function is given by

$$F(H) = z^{-1}(\beta)e^{-\beta H},$$

where $\beta$ is the inverse temperature. Beck and Cohen [15] assumed that this quantity fluctuates adiabatically slowly, namely that the time scale is much larger than the relaxation time for reaching local equilibrium. In that case, the distribution function of the non-equilibrium system consists in Boltzmann factors $\exp(-\beta H)$ that are averaged over the various fluctuating inverse temperatures

$$F(H) = \int_0^\infty g(\beta) z^{-1}(\beta) e^{-\beta H} d\beta,$$

where $z^{-1}(\beta)$ is a normalizing constant, and $g(\beta)$ is the probability distribution of $\beta$. Let us stress that $\beta^{-1}$ is a local variance parameter of a suitable observable, the Hamiltonian of the complex system in this case. Ordinary statistical mechanics are recovered in the limit $g(\beta) \to \delta(\beta-\beta')$. In contrast, different choices for the statistics of may lead to a large variety of probability distributions $F(H)$. Several forms for $g(\beta)$ have been studied in the literature, e.g. [15, 16]. Beck et al. [16] have argued that typical experimental data are described by one of three superstatistical universality classes, namely, $\chi^2$, inverse $\chi^2$, or log-normal superstatistics. The first is the appropriate one if η has contributions from ν Gaussian random variables $X_1, \ldots, X_\nu$ due to various relevant degrees of freedom in the system. As mentioned before η needs to be positive; this is achieved by squaring these Gaussian random variables. Hence, $\eta = \sum X_i^2$ is $\chi^2$ distributed with degree ν,

$$f(\eta) = \frac{1}{\Gamma(\nu/2)} \left(\frac{\nu}{2\eta_0}\right)^{\nu/2} \eta^{\nu/2-1} e^{-\nu\eta/2\eta_0} \quad (1)$$

The average of η is $\eta_0 = \int \eta f(\eta) d\eta$. The same considerations are applicable if $\eta^{-1}$, rather than η, is the sum of several squared Gaussian random variables. The resulting distribution $f(\eta)$ is the inverse $\chi^2$ distribution given by

$$f(\eta) = \frac{\eta_0}{\Gamma(\nu/2)} \left(\frac{\nu\eta_0}{2}\right)^{\nu/2} \eta^{-\nu/2-2} e^{-\nu\eta_0/2\eta} \qquad (2)$$

where again $\eta_0$ is the average of $\eta$. Instead of being a sum of many contributions, the random variable $\eta$ may be generated by multiplicative random processes. Then $\ln\eta = \sum \ln X_i$ is a sum of Gaussian random variables. Thus it is log-normally distributed,

$$f(\eta) = \frac{1}{\sqrt{2\pi}\nu\eta} e^{-[\ln(\eta/\mu)]^2/2\nu^2} \qquad (3)$$

which has an average $\mu\sqrt{w}$ and variance $\mu^2 w(w-1)$, where $w = \exp(\nu^2)$.

## RMT within superstatistics

The assumptions of RMT stated in the introduction lead to the following joint probability distribution function for the matrix elements in a random-matrix ensemble

$$P(\mathrm{H}) = Z^{-1}(\eta) e^{-\eta \mathrm{Tr}(\mathrm{H}^\dagger \mathrm{H})},$$

where $\eta$ is a parameter related the mean level density and $Z^{-1}(\eta)$ is a normalizing constant. To apply the concept of superstatistics to RMT, assumes the spectrum of a (mixed) system as made up of many smaller cells that are temporarily in a chaotic phase. Each cell is large enough to obey the statistical requirements of RMT but is associated with a different distribution of the parameter $\eta$ according to a probability density $f(\eta)$. Consequently, the superstatistical random-matrix ensemble used for the description of a mixed system consists of a superposition of Gaussian ensembles. Its joint probability density distribution of the matrix elements is obtained by integrating the distribution of the random-matrix ensemble over all positive values of $\eta$ with a statistical weight $f(\eta)$,

$$P(\mathrm{H}) = \int_0^\infty f(\eta) Z^{-1}(\eta) e^{-\eta \mathrm{Tr}(\mathrm{H}^\dagger \mathrm{H})} d\eta. \qquad (4)$$

Despite the fact that it is hard to make this picture rigorous, there is indeed a representation which comes close to this idea [17].

The new framework of RMT provided by superstatistics should now be clear. The local mean spacing is no longer uniformly set to unity but allowed to take different (random) values at different parts of the spectrum. The parameter $\eta$ is no longer a fixed parameter but it is a stochastic variable with probability distribution $f(\eta)$. Instead, the observed mean level spacing is just the expectation value of the local ones. The fluctuation of the local mean spacing is due to the correlation of the matrix elements which disappears for chaotic systems. In the absence of these fluctuations, $f(\eta) = \delta(\eta - \eta_0)$ and we obtain the standard RMT. Within the superstatistics framework, we can express any statistic $\sigma(E)$ of a mixed system that can in principle be obtained from the joint eigenvalue distribution by integration over some of the eigenvalues, in terms of the corresponding statistic $\sigma^{(G)}(E,\eta)$ for a Gaussian random ensemble. The superstatistical generalization is given by

$$\sigma(\mathrm{E}) = \int_0^\infty f(\eta) \sigma^{(G)}(\mathrm{E},\eta) d\eta. \qquad (5)$$

The remaining task of superstatistics is the computation of the distribution $f(\eta)$. The time series analysis in Ref. [18] allows to derive a parameter distribution $f(\eta)$, as we shall show now.

## Time-series representation

In this section, we use the time series method for the study of the fluctuations of the resonance spectra of mixed microwave billiards. Representing energy levels of a quantum system as a discrete time series has been probed in a number of recent publications (e.g., [19]). Billiards are often used as simple models in the study of Hamiltonian systems. A billiard consists of a point particle which is confined to a container of some shape and reflected elastically on impact with the boundary. The shape determines whether the dynamics inside the billiard is regular, chaotic or mixed. The best-known examples of chaotic billiards are the Sinai billiard (a square table with a circular barrier at its center) and the Bunimovich stadium (a rectangle with two circular caps) [20]. Neighboring parallel orbits diverge when they collide with dispersing components of the billiard boundary. Elliptic and rectangular billiards are examples of regular billiards. The Quantum-Chaos group in Darmstadt University carried out a series of experiments with microwave billiards [18, 21]. Here we summarize the time-series analysis of the resonance spectra of the so-called Limaçon billiard.

The "time-series" analysis of the spectra of billiards of both families manifests the existence of two relaxation lengths in the spectra of mixed systems, a short one defined as the average length over which energy fluctuations are correlated, and a long one that characterizes the typical linear size of the heterogeneous domains of the total spectrum. This is done in an attempt to clarify the physical origin of the heterogeneity of the matrix-element space, which justifies the superstatistical approach to RMT. The second main result of this section is to derive a parameter distribution f($\eta$)

**Limaçon billiard**
The Limaçon billiard is a closed billiard whose boundary is defined by the quadratic conformal map of the unit circle $z$ to $w$, $w = z + \lambda z^2$, $|z|=1$. The shape of the billiard is controlled by a single parameter $\lambda$ with $\lambda = 0$ corresponding to the (regular) circle and $\lambda = 1/2$ to the cardioid billiard, which is chaotic. For $0 \leq \lambda < 1/4$, the Limaçon billiard has a continuous and convex boundary with a strictly positive curvature and a collection of caustics near the boundary [22]. At $\lambda=1/4$, the boundary has zero curvature at its point of intersection with the negative real axis, which turns into a discontinuity for $\lambda > 1/4$. Accordingly, there the caustics no longer persist [23]. The classical dynamics of this system and the corresponding quantum billiard have been extensively investigated by Robnik and collaborators [24]. They concluded that the dynamics in the Limaçon billiard undergoes a smooth transition from integrable motion at $\lambda = 0$ via a soft chaos KAM regime for $0 < \lambda \leq 1/4$ to a strongly chaotic dynamics for $\lambda = 1/2$.

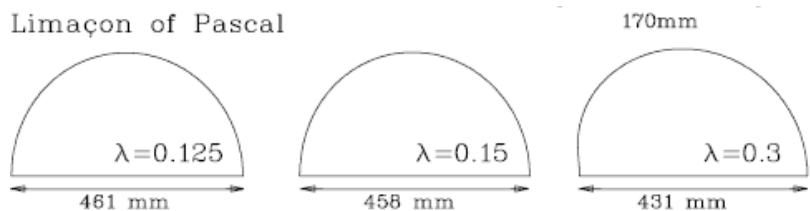

Fig. 1. Schematic represent of the billiards subject to the Darmstadt experiment

In the Darmstadt experiment, reported in [18], three de-symmetrized cavities with the shape of billiards from the family of Limaçon billiards (see Fig. 1). They have been constructed for the values $\lambda = 0.125, 0.150, 0.300$ and the first 1163, 1173 and 942 eigenvalues were measured, respectively. The latter billiard is of chaotic while the other two have mixed dynamics. More details on these experiments are given in [25].

**Two spectral-correlation lengths**
The results of the previous subsection confirm the assumption that the spectrum consists of a succession of cells with different mean level density. Each cell is associated with a relaxation length $\tau$, which is defined as that length-scale over which energy fluctuations are correlated. Our basic assumption is that the level sequence within each cell is modeled by a random-matrix ensemble. The relaxation length τ may also be regarded as an operational definition for the average energy separation between levels due to level repulsion. In the long-term run, the stationary distributions of this inhomogeneous spectrum arise as a superposition of the "Boltzmann factors" of the standard RMT, i.e. $e^{-\eta \text{Tr}(H^\dagger H)}$. The parameter $\eta$ is approximately constant in each cell for an eigenvalue interval of length $T$ (see Fig. 2). In superstatistics this superposition is performed by weighting the stationary distribution of each cell with the probability density $f(\eta)$ to observe some value $\eta$ in a randomly chosen cell and integrating over $\eta$. Of course, a necessary condition for a superstatistical description to make sense is the condition $\tau \ll T$, because otherwise the system is not able to reach local equilibrium before the next change takes place.

*The long time scale*: First, let us determine the long time scale $T$. For this we divide the level-spacings series into $N$ equal level-number intervals of size $n$. The total length of the spectrum is $Nn$. We then define the mean local kurtosis $\kappa(n)$ of a spacing interval of length $n$ by

$$\kappa(n) = \frac{1}{N} \sum_{i=1}^{N} \frac{\left\langle (s-\bar{s})^4 \right\rangle_{i,N}}{\left\langle (s-\bar{s})^2 \right\rangle^2_{i,N}}.$$

Here $\left\langle ... \right\rangle_{i,T} = \sum_{k=(i-1)n+1}^{in} ...$ denotes a summation over an interval of length $n$, starting at level spacing $in$, and $\bar{s}$ is either the local average spacing in each spacing interval or the global average $\bar{s} = 1$ over the entire spacings series. We chose the latter one. In probability theory and statistics, kurtosis is a measure for the "flatness" of the probability distribution of a real-valued random variable. Higher kurtosis means that a larger part of the contributions to the variance is due to infrequent extreme deviations, as opposed to frequent modestly sized ones. A superposition of local Gaussians with local flatness three results in a kurtosis $\kappa = 3$. We define the superstatistical level-number scale $T$ by the condition

$$\kappa(T) = 3, \qquad (6)$$

that is, we look for the simplest superstatistics, a superposition of local Gaussians [16]. If $n$ is chosen such, that only one value of $s$ is contained in each interval, then of course $\kappa(1) = 1$. If on the other hand $n$ comprises the entire spacing series, then we obtain the flatness of the distribution of the entire signal, which will be larger than 3, since superstatistical distributions are fat-tailed. Therefore, there exists a level-number scale $T$ which solves Eq. (6). Figure 2 shows the dependence of the local flatness of a spacing interval on its length for the two mushroom and the three Limaçon billiards. In the case of the chaotic Limaçon billiard, in which $\lambda=0.300$, the quantity $\kappa$ does not cross the line of $\kappa=3$ for the considered values of $n$. It is expected that $T = N$ in this case since the fluctuations in a chaotic (unfolded) spectrum are uniform. The values of $T$ for the mixed billiards with $\lambda = 0.125$ and $0.150$ are $12.5$ and $15.4$ respectively. For the chaotic billiard $\lambda = 0.300$, the curve of $\kappa(n)$ and the line $\kappa =3$ do not intersect as expected.

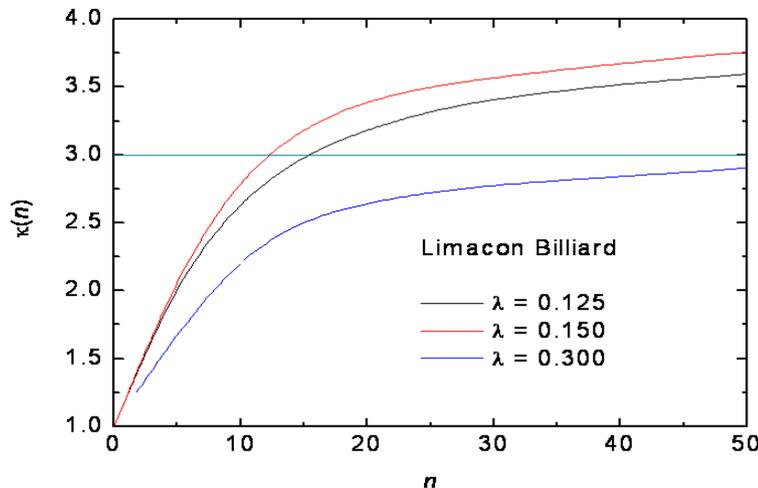

Fig. 2. Determination of the long correlation length $T$ from Eq. 6 as the point of intersection of the curve of $\kappa(n)$ with the line $\kappa = 3$.

*The short time scale:* The relaxation time associated with each of the $N$ intervals, was estimated in [16] from the small-argument exponential decay of the autocorrelation function

$$C(n)=(<s(i)s(i + n)>-1)/(s^2-1))$$

of the time series under consideration. Figure 3 shows the behavior of the autocorrelation functions for the series of resonance-spacings of the two families of billiards.

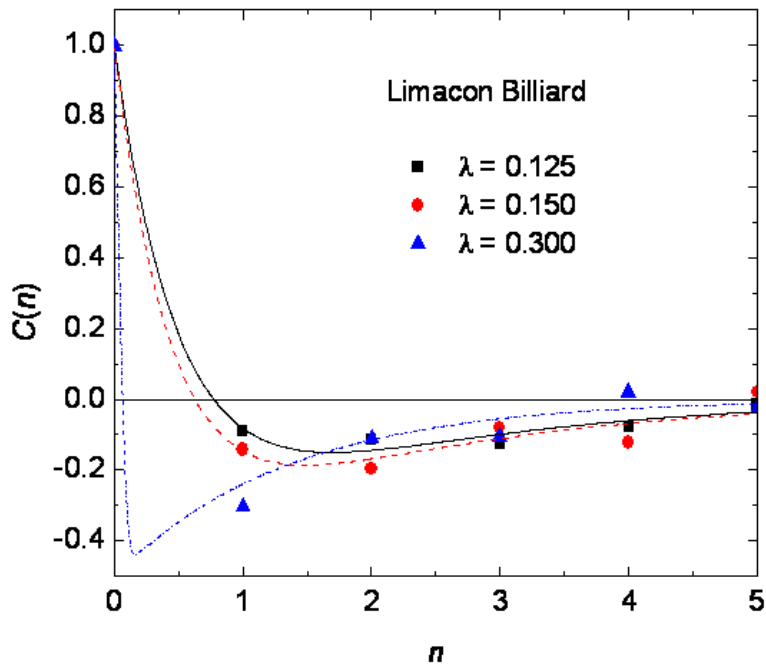

Fig. 3. Determination of the short correlation length $\tau$ from the autocorrelation function $C(n)$.

Quite frequently, the autocorrelation function shows single-exponential decays, $C(n) = \exp(-n/\tau)$, where $\tau>0$ defines a relaxation "time". A typical example is the velocity correlation of Brownian motion [26]. The autocorrelation functions studied here clearly do not follow this trend. For the systems with mixed dynamics, they decay rapidly from a value of $C(0) = 1$, change sign at some $n$ becoming negative, then asymptotically tend to zero. In an attempt to quantify the dependence of $C$ on $n$, we parameterized its empirical value in the form of a superposition of two exponentially decaying functions

$$C(n)=A_1\exp(-n/\tau_1) +A_2\exp(-n/\tau_2)$$

and (arbitrarily) fixed the superposition coefficient as $A_1 = 1.5$ and $A_2 = -0.5$. The curves in Fig. 3 show the resulting parameterization. The best fit parameters are $(\tau_1, \tau_2) = (0.51, 1.9)$,

(0.44, 2.1) and (0.2, 1.36) for. We may estimate τ as the mean values of $\tau_1$ and $\tau_2$ and conclude that τ has a value slightly larger than 1 for each billiard. This is sufficient to conclude that the ratio $T/\tau$ is large enough in each billiard to claim two well separated "time" scales in the level-spacings series, which justifies describing them within the framework of superstatistics.

**Estimation of the parameter distribution**
We represent the spectra of the three Limaçon billiards as discrete time series, in which the role of time is played by the level ordering. The distribution $f(\eta)$ is determined by the dynamics of the entire time series representing the spectrum of each billiard. Next, we need to determine which of these distributions fits best that of the slowly varying stochastic process $\eta(n)$ described by the experimental data. Since the variance of superimposed local Gaussians is given by $\eta^{-1}$, we may determine the process $\eta(t)$ from the series

$$\eta_i = \frac{1}{\langle s^2 \rangle_{i,T} - \langle s \rangle_{i,T}^2},$$

where $\langle ... \rangle_{i,T} = \sum_{k=(i-1)n+1}^{in} ...$ denotes a summation over an interval of length $n$, starting at level spacing $in$. The result for the $\lambda = 1/2$ billiard is shown below.

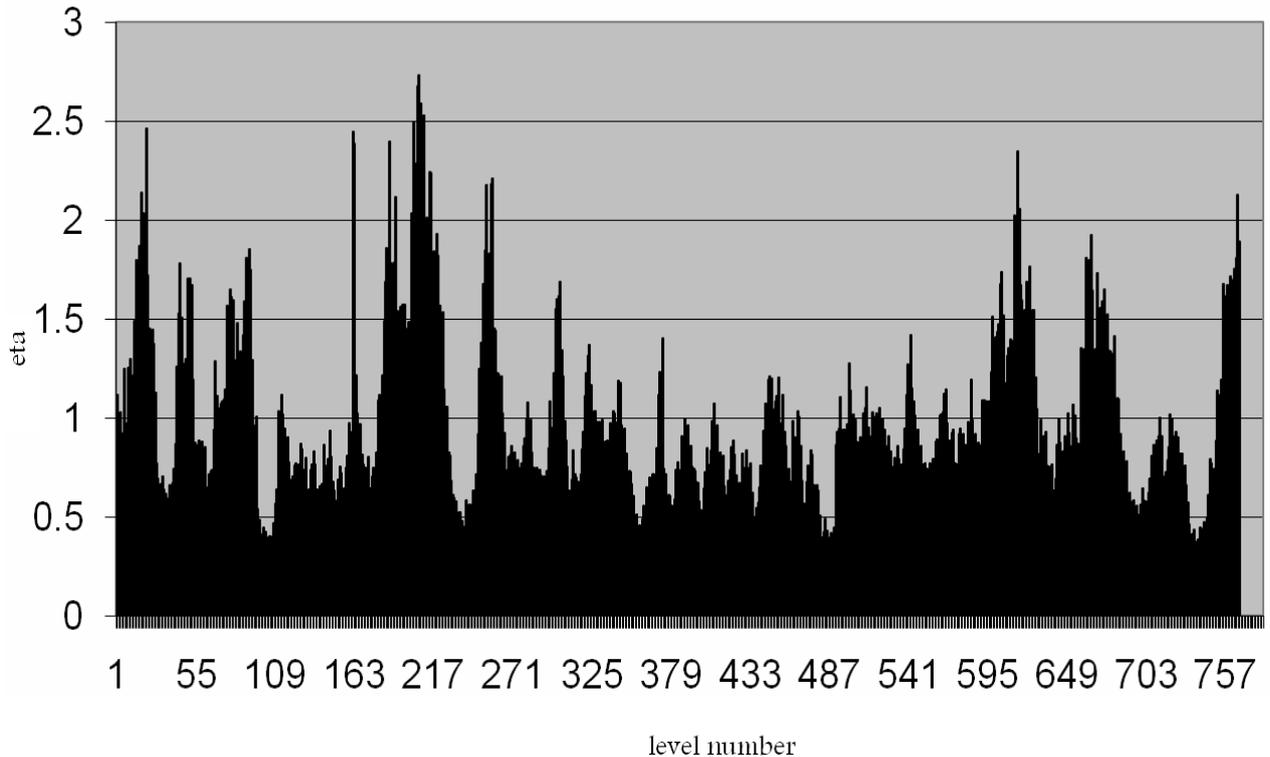

Fig. 2. The values extracted from the time series for the superstatistical parameter η.

The figure suggests that the extracted value of $\eta$ (and thus the level density) shows rapid fluctuations superimposed over slower ones. This agrees with the picture described by the assumptions of the superstatistical RMT that the spectrum of the mixed system is composed of segments with different mean level density.

The probability density $f(\eta)$ is determined from the histogram of the $\eta(i)$ values for all $i$. The resulting experimental distributions are shown in Fig.3. We compared them with the log-normal, the $\chi^2$ and the inverse $\chi^2$ distributions with the same mean $\langle\eta\rangle$ and variance

$\langle\eta^2\rangle - \langle\eta\rangle^2$. The inverse $\chi^2$ distribution fits the data significantly better than the other two distributions.

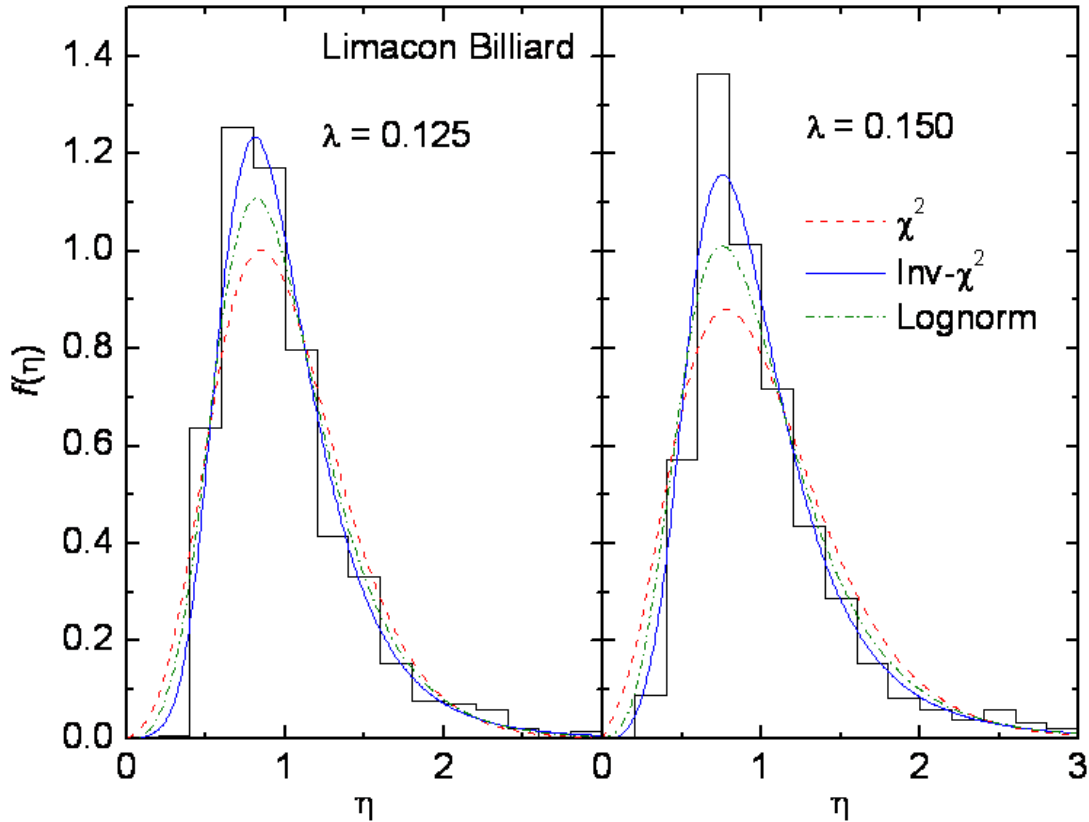

Fig. 3. The parameter distributions obtained from the time-series analysis of the mixed Limaçon billiards compared with the distributions in Eqs. (1-3).

**Nearest-neighbor spacing distribution**

This section focuses on the question whether the inverse $\chi^2$ distribution of the superstatistical parameter $\eta$ in Eq. (2) is suitable for describing the nearest-neighbor spacing distribution (NNSD) of systems in the transition out of chaos within the superstatistical approach to RMT. As mentioned above, the NNSD of a chaotic system is well described by that of random matrices from the GOE, which is well approximated the Wigner surmise, if the system is chaotic and by that for Poisson statistics if it is integrable. Numerous interpolation formulas describing the intermediate situation between integrability and chaos have been proposed [8]. One of the most popular NNSDs for mixed systems is elaborated by Berry and Robnik [27]. This distribution is based on the assumption that semi-classically the eigenfunctions are localized either in classically regular or chaotic regions in phase space. Accordingly, the sequences of eigenvalues connected with these regions are assumed to be statistically independent, and their mean spacing is determined by the invariant measure of the corresponding regions in phase space. The largest discrepancy between the empirical NNSDs of mixed systems and the Berry-Robnik (BR) distribution is observed for level spacings $s$ close to zero. While the formers almost vanish for $s = 0$, the latter approaches a constant and nonvanishing value for $s \rightarrow 0$.

It follows from Eq. (5) that the statistical measures of the eigenvalues of the superstatistical ensemble are obtained as an average of the corresponding $\eta$-dependent ones of standard RMT weighted with the parameter distribution $f(\eta)$. In particular, the superstatistical NNSD is given by

$$p(s) = \int_0^\infty f(\eta) p_w(\eta, s) d\eta \qquad (7)$$

where $p_w(\eta, s)$ is the Wigner surmise for the Gaussian orthogonal ensemble with the mean spacing depending on the parameter $\eta$,

$$p_w(\eta, s) = \eta s \exp\left(\frac{1}{2}\eta s^2\right). \qquad (8)$$

For a $\chi^2$ distribution of the superstatistical parameter $\eta$, one substitutes Eqs. (1) and (8) into Eq. (7) and integrates over $\eta$. The resulting NNSD is given by

$$p_{\chi^2}(\nu, s) = \frac{\eta_0}{\left(1 + \eta_0 s^2 / \nu\right)^{1+\nu/2}} \qquad (9)$$

The parameter $\eta_0$ is fixed by requiring that the mean-level spacing $\langle s \rangle = \int_0^\infty s p(s) ds$ equals unity. For an inverse $\chi^2$ distribution of $\eta$, given by Eq. (2), one obtains the following superstatistical NNSD

$$p_{Inv\,\chi^2}(\nu, s) = \frac{2\eta_0 s}{\Gamma(\nu/2)} \left(\sqrt{\eta_0 \nu}\,\frac{s}{2}\right)^{\nu/2} K_{\nu/2}\left(\sqrt{\eta_0 \nu} s\right), \qquad (10)$$

where $K_m(x)$ is a modified Bessel function [28], $\Gamma(x)$ is a gamma function and $\eta_0$ again is determined by the requirement that the mean-level spacing $<s> = 1$. Finally, if the parameter $\eta$ has a normal distribution (3), then the NNSD for this distribution,

$$p_{\text{log-norm}}(s) = \int_0^\infty \frac{1}{\sqrt{2\pi}\nu\eta} e^{-[\ln(\eta/\mu)]^2/2\nu^2} p_w(\eta, s) d\eta, \qquad (11)$$

cannot be evaluated analytical and has to be calculated numerically.

We have compared the resulting NNSDs given in Eqs. (9), (10) and (11) with the experimental ones for the two Limaçon billiards with mixed dynamics. In Fig. 4 the experimental results are shown together with the superstatistical and the BR distribution [27].

**Summary**

Superstatistics has been applied to study a wide range of phenomena, ranging from turbulence to econophysics. RMT is among these. In the later application, the parameter distributions of the random-matrix ensembles have been obtained by assuming suitable forms or applying the principle of maximum entropy. In this paper we use the time-series method to show that the spectra of mixed systems have two correlation scales as required for the validity of the superstatistical approach. The time-series analysis also shows that the best choice of the superstatistical parameter distribution for a mixed system is an inverse $\chi^2$ distribution. We calculate the corresponding NNS distribution of the energy levels and compare it with the spectrum of two microwave resonators of mushroom-shaped boundaries and two of the family of Limaçon billiards, which exhibit mixed regular-chaotic dynamics. Resonance-strength distributions for the Limaçon billiards are also analyzed. In all cases the experimental data are found in better agreement with the corresponding distributions with inverse-$\chi^2$ superstatistics than all the other considered distributions including the celebrated Brody NNS distribution and the resonance-strength distribution that follows from the Alhassid-Novoselsky Gamma distribution of transition intensities.

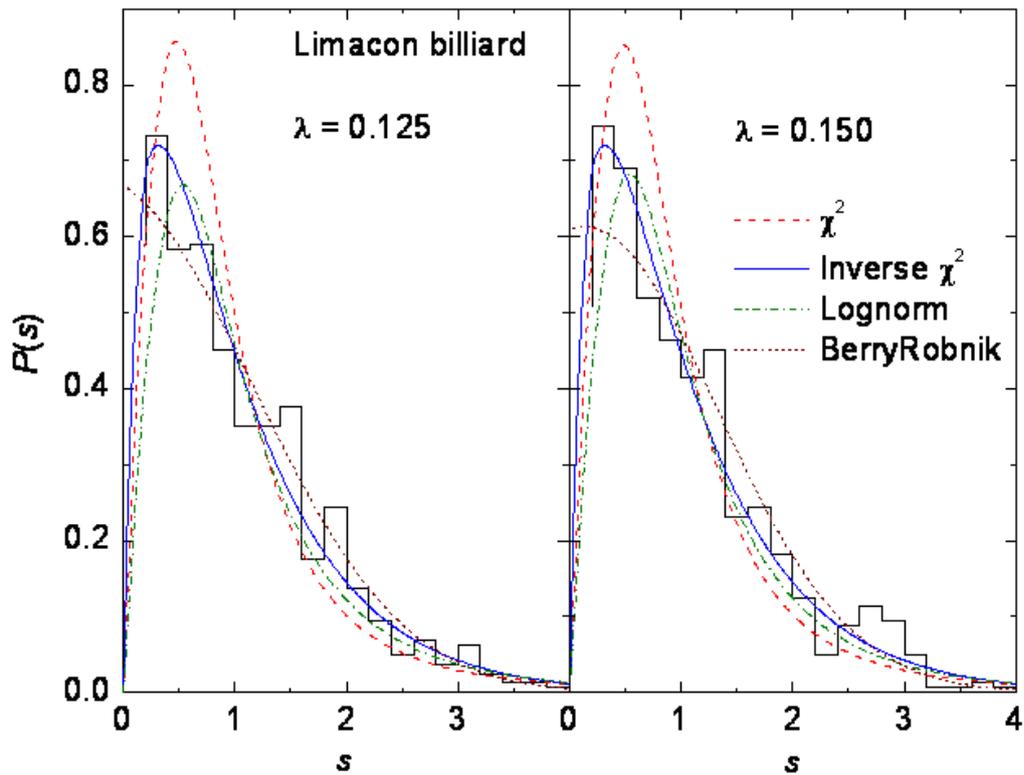

Fig. 4. Experimental NNS distributions for the two mixed Limaçon billiards compared with the superstatistical distributions. The solid lines are for the inverse $\chi^2$ distribution, the dashed lines for the $\chi^2$ distributions, the dashed-dotted lines for the log-normal distribution and the short-dashed line for the BR distribution.

# References

[1] A.J. Lichtenberg, M.A. Lieberman, Regular and Stochastic Motion, Applied Mathematical Sciences (Springer, New York, 1983).
[2] J.-P. Eckmann and D. Ruelle, Rev. Mod. Phys. 67, 617 (1985).
[3] S. S. E. H. Elnashaie and S. S. Elshishini, Dynamical Modelling, Bifurcation and Chaotic Behavior of Gas-Solid Catalytic Reactions (Gordon and Breach, Amsterdam, 1996).
[4] L. A. Bunimovich and S. Venkatuyiri, Phys. Rep. 290, 81 (1997).
[5] M.V. Berry , J. Phys. A 10, 2083 (1977).
[6] O. Bohigas, M.J. Giannoni, and C. Schmit, Phys. Rev. Lett. 52, 1 (1984).
[7] M.L. Mehta, Random Matrices 2nd ed. ( Academic, New York, 1991).
[8] T. Guhr, A. Müller-Groeling, and H. A. Weidenmüller, Phys. Rep. 299, 189 (1998).
[9] R. Balian, Nuovo Cim. 57, 183 (1958).
[10] N. Rosenzweig and C. E. Porter, Phys. Rev. 120, 1698 (1960).
[11] M. S. Hussein and M. P. Pato, Phys. Rev. Lett. 70, 1089 (1993); Phys. Rev. C 47, 2401 (1993).
[12] A. Casati, L. Molinari, and F. Izrailev, Phys. Rev. Lett. 64, 1851 (1990); Y. V. Fyodorov and A. D. Mirlin, Phys. Rev. Lett. 67, 2405 (1991); A.D. Mirlin, Y.V. Fyodorov, F.M. Dittes, J. Quezada, and T.H. Seligman, Phys. Rev. E 54, 3221 (1996); V.E. Kravtsov, K.A. Muttalib, Phys. Rev. Lett. 79, 1913 (1997); F. Evers and A.D. Mirlin, Phys. Rev. Lett. 84 3690 (2000); Phys. Rev. B 62, 7920 (2000).
[13] J. Evans and F. Michael, e-prints cond-mat/0207472 and /0208151;  F. Toscano, R. O. Vallejos, and C. Tsallis, Phys. Rev. E 69, 066131 (2004) ;F. D. Nobre and A.M. C. Souza, Physica A 339, 354 (2004); A. Y. Abul-Magd, Phys. Lett. A 333, 16 (2004); A. C. Bertuola, O Bohigas, and M. P. Pato, Phys. Rev. E 70, 065102(R) (2004); A.Y. Abul-Magd, Phys. Rev. E 71, 066207 (2005); A.Y. Abul-Magd, Phys. Lett. A 361, 450 (2007).
[14] A.Y. Abul-Magd, Physica A 361, 41 (2006); A.Y. Abul-Magd, Phys. Rev. E 72, 066114 (2005).
[15] C. Beck and E. G. D. Cohen, Physica A 322, 267 (2003); E. G. D. Cohen, Physica D 193, 35 (2004); C. Beck, Physica D 193, 195 (2004); C. Beck, Europhys. Lett. 64, 151 (2003); F. Sattin and L. Salasnich, Phys. Rev. E 65, 035106(R) (2003); F. Sattin, Phys. Rev. E 68, 032102 (2003); A. Reynolds, Phys. Rev. Lett. 91, 084503 (2003); M. Ausloos and K. Ivanova, Phys. Rev. E 68, 046122 (2003); C. Beck, Physica A 331, 173 (2004).
[16] C. Beck, E. G. D. Cohen, and H. L. Swinney, Phys. Rev. E 72, 056133 (2005).
[17] G. Le Caër and R. Delannay, Phys. Rev. E 59, 6281 (1999); K. A. Muttalib and J. R. Klauder, Phys. Rev. E 71, 055101(R) (2005).
[18] A.Y. Abul-Magd, B. Dietz, T. Friedrich, and A. Richter, Phys. Rev. E 77, 046202 (2008).
[19] A. Relaño, J. M. G. Gómez, R. A. Molina, J. Retamosa, and E. Faleiro, Phys. Rev. Lett. 89, 244102 (2002); E. Faleiro, J. M. G. Gómez, R. A. Molina, L. Muñoz, A. Relaño, and J. Retamosa, Phys. Rev. Lett. 93, 244101 (2004); E. Faleiro, U. Kuhl, R. A. Molina, L. Muñoz, A. Relaño, and J. Retamosa, Phys. Lett. A 358, 251 (2006); R.A. Molina, J. Retamosa, L. Muñoz, A. Relaño, and E. Faleiro, Phys. Lett. B 644, 25 (2007).
[20] L. Bunimovich, Funct. Anal. Appl. 8, 254 (1974).
[21] B. Dietz, T. Friedrich, M. Miski-Oglu, A. Richter, T. H. Seligman, and K. Zapfe, Phys. Rev. E 74, 056207 (2006); B. Dietz, T. Friedrich, M. Miski-Oglu, A. Richter, and F. Schäfer, Phys. Rev. E 75, 035203(R) (2007).
[22] M. Robnik, J. Phys. A 16, 3971 (1983); R. Dullin and A. Bäcker, Nonlinearity 14, 1673 (2001).
[23] M. Gutiérrez, M. Brack, K. Richter, and A. Sugita, J. Phys. A: Math. Gen. 40, 1525 (2007).

[24] T. Prosen and M. Robnik, J. Phys. A 27, 8059 (1994); B. Li and M. Robnik, J. Phys. A 28, 2799 (1995).

[25] C. Dembowski, H.-D. Gräf, A. Heine, T. Hesse, H. Rehfeld, and A. Richter, Phys. Rev. Lett. 86, 3284 (2001); H. Rehfeld, H. Alt, H.-D. Gräf, R. Hofferbert, H. Lengeler, and A. Richter, Nonlinear Phenomena in Complex Systems 2, 44 (1999).

[26] F. Reif, Fundamental of Statistical and Thermal Physics (McGrow-Hill, Singapore, 1985).

[27] M. V. Berry and M. Robnik, J. Phys. A 17, 2413 (1984).

[28] I. S. Gradshteyn and I. M. Ryzhik, Tables of Integrals, Series, and Products (Academic, New York, 1980).